\documentclass[pra,bibnotes,twocolumn]{revtex4}
\usepackage{graphicx}
\usepackage{hyperref}
\usepackage{graphicx}
\usepackage{amsmath,amssymb}
\usepackage{float}
\usepackage{bm}

\begin{document}
\draft
\def\ds{\displaystyle}
\title{Quasi-stationary solutions in non-Hermitian systems}
\author{Cem Yuce}
\email{cyuce@eskisehir.edu.tr}
\address{Department of Physics, Faculty of Science, Eskisehir Technical University, Turkey }

\date{\today}
\begin{abstract}
Eigenstates exhibit localization at an open edge in a non-Hermitian lattice due to non-Hermitian skin effect. We here explore another interesting feature of non-Hermitian skin effect and predict quasi-stationary solutions, which are approximately time-independent. We show that the transition from such states to eigenstates is dramatically non-perturbative. We discuss that mathematically extending the boundary of long non-Hermitian lattices to infinity can lead to nontrivial solution. We consider a non-Hermitian variant of the Su-Schrieffer-Heeger (SSH) model and predict non-topological but robust quasi-stationary zero energy modes.
\end{abstract}
\maketitle

\section{Introduction}

Non-Hermitan systems are ubiquitous in nature and has attracted great attention in the last two decades. There are many ongoing experiments to check theoretical predictions for non-Hermitian systems. Photonic lattices have become the most popular area in the last decade in the community of non-Hermitian physics. Introducing gain and loss to a photonics lattice is not the only way to obtain a non-Hermitian lattice. Anisotropy of hopping amplitudes between adjacent sites leads also to non-Hermiticity. This way of obtaining non-Hermiticity is of particular interest as the spectrum strongly depends on the boundary conditions. In the case of open boundary conditions (OBC), all eigenstates are localized around the boundary and no extended modes in the bulk of the lattice exist. This unique behavior is dubbed non-Hermitian skin effect \cite{nhs1}, originally introduced in the study of topological phase of non-Hermitian systems \cite{reva1,reva2,reva3,reva4,reva420}. This extreme localization of all eigenstates implies that the bulk-boundary correspondence fails \cite{nhs3,nhs2,nhs4,nhs5,cemAnnals,nhs10,nhs11,nhs12}. Therefore, the theory of non-Hermitian topological insulating phase is more complex and richer predictions compared to its Hermitian counterpart \cite{nhs7,nhs8,nhs9,nhs14,nhs15,nhs16,nhs17,refek2,refek3,refek4,refek5,refek6,refek7,refek8,refek9,refek10,refek11,refek12,refek13,refek14}. The non-Hermitian skin effect has some interesting consequences such as light funneling \cite{funnel}, in which any light excitation in the lattice moves towards the open boundary and non-Bloch band theory of non-Hermitian systems \cite{nhs6,nhs13}. The non-Hermitian skin effect is not only theoretical curiosity as it has recently been realized in some experiments \cite{nhs18,nhs19,nhs20,nhs21}. Recently, we predict eigenstate clustering around exceptional points and show that eigenstate localization due to non-Hermitian skin effect is a special type of eigenstate clustering \cite{cluster}.  \\
In this paper, we predict another interesting consequence of non-Hermitian skin effect. We find that very long and infinitely long lattices can have dramatically different spectra as opposed to Hermitian systems. We explore quasi-stationary states with continuum spectrum in a finite lattice with open boundary conditions. They are not truly stationary but stay stationary in the time scale of a typical experiment in photonics systems \cite{ourpaperc1,ourpaperc2}. We discuss that the transition from the eigenstates to quasi-stationary states is dramatically non-perturbative. These unique states can be used in a variety of applications in non-Hermitian systems. We apply our idea to a one dimensional non-Hermitian topological system and predict non-topological but robust quasi-stationary zero energy modes.

\section{Quasi-stationary solutions}

Consider a non-Hermitian tight-binding lattice with asymmetric hopping. The time-independent complex field amplitude $\ds{  \psi_j }$ at site $j$ is governed
\begin{equation}\label{8291bdool} 
\psi_{j+1}   + {\gamma}~\psi_{j-1}   =E ~ \psi_{j}  
\end{equation} 
where $\ds{ j=1,2,...,N }$ and $\ds{N}$ is the total number of lattice sites, $\ds{ 0\leq \gamma<1 }$ is the non-Hermitian degree and $\ds{ E  }$ are energy eigenvalues. This equation is a second order linear homogeneous recurrence relation and its most general solution is given by
 \begin{equation}\label{co008ow} 
\psi_{j} =   \Delta_{+}^j  ~ c_1 + \Delta_{-}^j ~c_2  
\end{equation} 
where  
 $\ds{ \Delta_{\mp}= \frac{E \mp\sqrt{E^2-4\gamma}}{2}   }$ and $\ds{c_{1,2}}$ are arbitrary constants. The energy eigenvalues and the arbitrary constants $\ds{c_{1,2}}$ are determined from the boundary conditions and normalization condition. Applying OBC to the solution (\ref{co008ow}), $\ds{\psi_0=\psi_{N+1}=0}$, yields $\ds{c_2=-  c_1}$ and
\begin{equation}\label{ccywop201a} 
 \Delta_{+}^{N+1}-  \Delta_{-}^{N+1}=0 
\end{equation} 
This equation can readily be solved: $\ds{   \Delta_{+} = e^{i\frac{2n\pi}{N+1}} ~\Delta_{-}  }$, where $\ds{n=1,..,N}$. We then obtain the corresponding energy eigenvalues for OBC
\begin{equation}\label{c5793els} 
E_n=2~\sqrt{ \gamma}  \cos  \frac{n \pi}{N+1} 
\end{equation} 
Note that the energy eigenvalues for periodic boundary condition (PBC) are given by $\ds{   E_{k}=e^{ik  }+{\gamma}~e^{-ik  }   }$ corresponding to the extended eigenstates $\ds{  \psi_{j}=\psi_0~ e^{ikj}  }$. Using $\ds{\psi_{N+1}=  \psi_1}$, we find $\ds{k= \frac{2 m \pi}{N} }$ and $\ds{ m=0,1,...,N-1}$. Let us now compare the spectra for OBC and PBC. Firstly, the eigenvalues are real for OBC while they are complex for PBC. Secondly, an exceptional point occur for OBC at $\ds{\gamma=0}$ since all eigenstates coalesce to a localized exceptional eigenstate with zero energy, $\ds{\psi_j=\psi_1~\delta_{j,1}}$. However, no exceptional points occurs for PBC. Such differences are especially interesting if we consider a long lattice with a small value of $\ds{\gamma}$. In fact, the PBC lattice has one additional bond between the first and last lattice sites compared to the OBC lattice, but such a slight difference has nonlocal and non-perturbative consequences on both eigenstates and eigenvalues. This is known as non-Hermitian skin effect.\\
We have derived Eq. (\ref{c5793els}) by assuming that $\ds{N}$ is a finite number. Surprisingly, there is another solution of (\ref{ccywop201a}) only when $\ds{ N\rightarrow \infty}$ (semi-infinite lattice for which energy eigenvalues form a continuum band). The solution reads $\ds{\Delta_{\mp}<1}$ for which the OBC is also satisfied, $\ds{ \psi_{\infty}  =0}$. If we solve the equation $\ds{\Delta_{\mp}<1}$, we obtain a continuum band inside an ellipse in the complex plane
\begin{equation}\label{c57isbf04s} 
\frac{E_R^2}{(1+\gamma)^2}+\frac{E_I^2}{(1-\gamma)^2}<1
\end{equation} 
where $\ds{E_R}$ and $\ds{E_I}$ are real and imaginary parts of energy eigenvalues, respectively. Note that the PBC solutions for finite and semi-infinite lattices have the same form, but $\ds{k}$ takes continuous values for the latter. It is interesting to see that the PBC spectrum formula $\ds{   E_{k}=e^{ik  }+{\gamma}~e^{-ik  }   }$ is equivalent to $\ds{\frac{E_R^2}{(1+\gamma)^2}+\frac{E_I^2}{(1-\gamma)^2}=1}$. This means that the PBC energy spectrum makes a closed loop enclosing the OBC spectrum. We plot them in Fig.1 for $\ds{\gamma=0}$ and $\ds{\gamma=0.6}$. The light blue region is for OBC while the red curve is for PBC. The elliptic band is compressed by increasing $\ds{\gamma}$ and flattened to a line on the $E_R$ axis in the Hermitian limit $\ds{\gamma=0}$. This means that PBC and OBC spectra for the semi-infinite lattice overlap only for the Hermitian case.\\
Let us also compare (\ref{c5793els}) and (\ref{c57isbf04s}). The former one becomes $\ds{  -2\sqrt{ \gamma} <E<2 \sqrt{\gamma }}$ in the limit $\ds{ N\rightarrow \infty}$, which predicts a real valued continuum band in between $\ds{\mp2\sqrt{\gamma}}$. This band is quite different from what Eq. (\ref{c57isbf04s}) predicts. This shows that our solutions, which appear only for the infinitely long lattice are novel. Furthermore, this novel solution is unique to non-Hermitian systems. In the Hermitian case $\ds{\gamma=1}$, the expressions (\ref{c5793els}) and (\ref{c57isbf04s}) yield the same results: $\ds{ - 2<E<2  }$, which is also what PBC predicts since $\ds{   E_k=2 \cos {k}  }$. As a result, we say that mathematically extending the boundary of long non-Hermitian lattices to infinity may not be trivial. In Hermitian systems, such an extension is always trivial as it slightly changes the spectrum. However, the spectrum of a very long lattice can be dramatically different from the spectrum of the infinitely long lattice in non-Hermitian systems. We stress that one can not obtain these novel solutions directly by numerically solving the eigenvalue equation since the corresponding matrix is infinite dimensional. To this end, we mention another important difference between the finite and semi-infinite open lattices. The exceptional point at $\ds{\gamma=0}$, which exist for a finite lattice disappears for the semi-infinite lattice. In other words, there exist only one zero-energy solution $\ds{\psi_{j} = \psi_1 ~\delta_{j,1} }$ for a finite lattice at $\gamma=0$ while infinitely many solutions $\ds{\psi_{j} =  E^{j-1}  ~\psi_1 }$ with complex valued energy $\ds{|E|<1}$ for the semi-infinite lattice at $\ds{\gamma=0}$ ($\ds{\psi_{\infty}=0}$ when $\ds{|E|<1}$).\\
We conclude that we use (\ref{c5793els}) for the spectrum for a finite lattice no matter how long the lattice is, while we use (\ref{c57isbf04s}) for the spectrum for the semi-infinite lattice. The latter formula can't be obtained from the former one in the limit $\ds{ N\rightarrow \infty}$. A real physical system has a finite number of lattice sites, so only the solution (\ref{c5793els}) is physical. Fortunately, the solution (\ref{c57isbf04s}) can be utilized for a sufficiently long lattice. Assume that $N$ is large. Then the right open boundary condition is not exactly but approximately satisfied. For example, suppose $\ds{\gamma=0}$ and $\ds{N=100}$ at which the right open boundary condition is exactly satisfied only when $E=0$. However, if we use Eq. (\ref{co008ow}), we get $\ds{\psi_{101}\approx 3.9 \times 10^{-31}}$ at $\ds{E=0.5}$ and $\ds{\psi_{101}\approx 1.6\times 10^{-10}}$ at $\ds{E=0.8}$. These are very small numbers and these solutions can be considered as quasi-stationary solutions. They eventually change their forms since they are not true eigenstates. Fortunately, these quasi-stationary solutions stay almost stationary in the time scale of a typical non-Hermitian experiment with waveguide arrays \cite{fswg}. The transition from quasi-stationary states to the eigenstates is non-perturbative. This can be simply seen from the above example where the quasi-stationary states can have the energies $E=0.5$ and $E=0.8$, which are non-pertubatively different from the exact zero energy eigenvalue.\\
Let us now qualitatively discuss quasi-stationary solutions. They appear in a sufficiently long and highly nonreciprocal latices where non-Hermitian skin effect occurs. In such a lattice, all eigenstates exhibit localization at either edge. If the localization occurs at the left edge, then we say that all eigenstates have extremely low densities at the right edge of the sufficiently long lattice. The key idea here is that the right boundary condition is automatically satisfied to a very good approximation. This has an interesting implication. Lifting the constraint at the right boundary leads to continuum of energy values since the energy values don't need to take some certain values to satisfy the right boundary condition. They are not true eigenstates but quasi-stationary solutions as OBC is approximately satisfied thanks to the non-Hermitian skin effect. \\
Quasi-stationary solutions can also occur  for non-Hermitian systems with gain and loss provided that non-Hermitian skin effect occurs. Let us illustrate our idea on an another model for further understanding of quasi-stationary solutions. Consider a tight-binding non-Hermitian lattice with alternating gain and loss. The field amplitude satisfies 
\begin{equation}\label{0p91bdn?bol} 
\psi_{j+1} +\gamma~\psi_{j-1}  +i ~(-1)^j~V_0~\psi_{j}   =E ~ \psi_{j}  
\end{equation} 
where $\ds{i}$ is the imaginary number, $j=1,2,..,N$, and the real number $\ds{  V_0  }$ is the gain/loss strength. Since the lattice is open, $\ds{\psi_{0}  =\psi_{N+1} =0}$. \\
We start with the case $\ds{\gamma=0}$ at which analytical solutions are available. In this case, there are two exceptional points of order $N/2$ and the eigenstates coalesce to two eigenstates $ \ds{    \psi_{ j,-}  =\{ 1,0,0,...,0  \}^T      }$ and $ \ds{    \psi_{j, +}  =\frac{1}{\sqrt{1+4V_0^2}}\{ i,-2 V_0 ,0,...,0  \}^T      }$ corresponding to the eigenvalues $E_{\mp}={\mp}~i~V_0$.  Let us find the unnormalized novel solutions for the semi-infinite lattice recursively. Assume $\ds{\psi_1=1}$. Then $\ds{\psi_2=E+iV_0}$, $\ds{\psi_3=( E+iV_0 )  (E-iV_0 )}$ and so on. The field amplitude at infinity becomes
\begin{equation}\label{cys77} 
\psi_{ \infty  }  = \left(  (  E+iV_0 )(  E-iV_0 )\right)^{ \infty   }  
\end{equation} 
Note that this recursive solution is reduced to the eigenstates if we replace $E=E_{\mp}={\mp}iV_0$. To obtain the energy eigenvalues for the semi-infinite lattice, we require $\ds{   |E^2+V_0^2|<1   }$. This condition is satisfied if $E$ is inside a continuum band in the complex plane
\begin{equation}\label{qi4ujhfg?904s} 
4 E_R^2 E_I^2 + (E_R^2 - E_I^2 + V_0^2)^2<1
\end{equation} 
The continuum band is inside a circle of radius $1$ in complex plane at $V_0=0$. As $V_0$ is increased, the continuum band is elongated on $E_I$ axis until $V_0=1$ at which the band looks like the $8$ shape. For $V_0>1$, the band is divided into two daughter bands, well separated in  $E_I$ axis. To study whether PBC spectrum encloses the OBC bands, we perform numerical calculation for various values of $V_0$ for very large number of $N$. We numerically find that PBC spectrum satisfy $\ds{4 E_R^2 E_I^2 + (E_R^2 - E_I^2 + V_0^2)^2=1}$. This result is consistent with our statement. Note that the two exceptional point for a finite lattice disappear for the semi-infinitely long lattice.\\
We have so far studied the fully asymmetric hopping at which analytical solutions are available. Consider now the case with $\ds{\gamma\neq0}$. In this case, obtaining the spectrum analytically is challenging. Here, we follow another approach to find the continuum band. We start to obtain the PBC spectrum numerically for a large lattice (suppose $N=10^5$). The PBC spectrum is a closed loop in complex domain and we suppose that it determines the boundary for the continuum band. To check this statement, we arbitrarily choose many $E$ values inside or outside this closed loop. We make iterations for Eq. (\ref{0p91bdn?bol}) with the initial value $\ds{\psi_1=1}$ at these arbitrary values of $E$. We numerically see that $\ds{\psi_j}$ goes to zero as $j$ goes to $N$ if $E$ is inside the closed loop, while $\ds{\psi_j}$ diverges if $E$ is outside the closed loop. This proves our above statement that the PBC spectrum encloses the OBC band in complex domain.
\begin{figure}[t]\label{2678ik0}
\includegraphics[width=4.25cm]{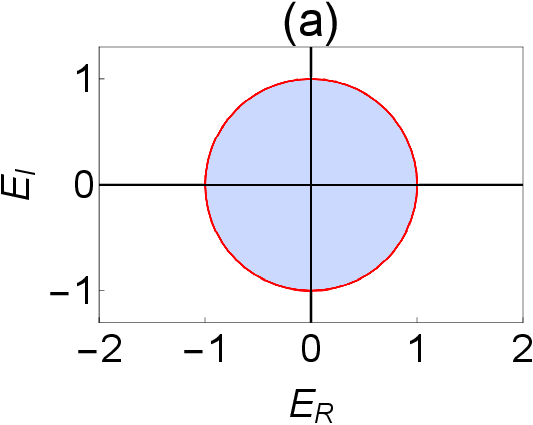}
\includegraphics[width=4.25cm]{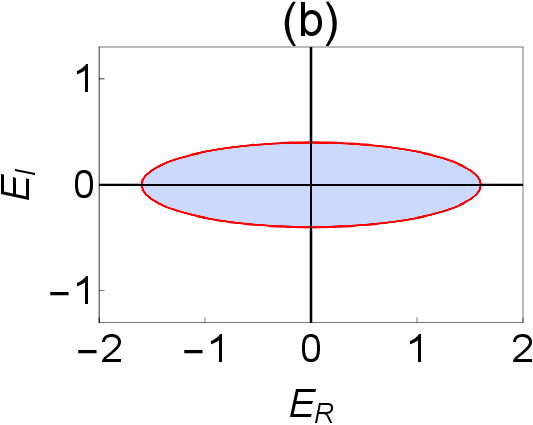}
\caption{ The plots show the elliptic continuum bands in the complex plane for the semi-infinite lattice at $\ds{ \gamma=0 }$ (a) and $\ds{ \gamma=0.6 }$ (b). The red loop is for PBC and the blue shaded area is for OBC. The PBC spectrum encloses the OBC spectrum. The band becomes a line on $E_R$ axis in the Hermitian limit $\ds{\gamma=1}$. In this case, the PBC and OBC spectra overlap.}
\end{figure}

\section{Non-topological robust zero-energy edge states}

We have discussed that quasi-stationary states for a finite lattice can be quite different from the eigenstates. Therefore quasi-stationary states bring us new degrees of freedom in non-Hermitian systems. Here we present an example where zero-energy quasi-stationary states appear in $\it{ topologically ~trivial}$ region of a 1D non-Hermitian system. We show that they are surprisingly immune to the same kind of disorder as topological zero-energy eigenstates.\\
Let us consider a non-reciprocal extension of the well-known Su-Schrieffer-Heeger (SSH) model \cite{reva1,reva2}, which is a one dimensional dimerized tight-binding lattice with non-reciprocal alternating hopping amplitudes. The Hamiltonian reads 
\begin{equation}\label{rof64oalk2} 
\hat{H}= \sum_{n=1}^{N-1}  T_n  ~  \left( ~  |n><n+1| +  \gamma~   |n+1><n|  ~\right)
\end{equation} 
where $\ds{T_n= 1+(-1)^n \delta   }$ are the alternating hopping amplitudes with $\ds{-1<\delta<1}$ and $\ds{0\leq\gamma<1}$ is the non-Hermitian degree and the total number of lattice sites $\ds{N}$ is assumed to be an even number. Note that non-Hermitian skin effect occurs and all eigenstate become localized at the left edge when $\ds{ \gamma }$ is a small number.\\
In the Hermitian system, $\ds{\gamma=1}$, it is well known that the point $\ds{\delta=0}$ is the topological phase transition point and topological zero energy modes appear when $\ds{\delta>0}$. In the non-Hermitian case, two topological zero-energy modes are still available. However, both of them as well as bulk eigenstates are localized around the left edge due to the non-Hermitian skin effect. In our recent paper \cite{cemAnnals}, we showed that a zero-energy mode at the right edge can still be available, but it tends to move rapidly to the left edge.\\
One can obtain zero-energy quasi-stationary states by solving the corresponding recursion equation for the above Hamiltonian. They are given by
\begin{equation}\label{8i9eoadskjcy} 
\psi_{E=0}=A   \sum_{n=1}^{ N/2} \left(\frac{-\gamma~(1-\delta)}{~~~~(1+\delta)}  \right)^{n-1}    |2n-1>
\end{equation} 
where $\ds{  A }$ is the normalization constant and the density is non-zero only for odd numbers of lattice sites. This zero energy solution satisfies the OBC for the semi-infinite lattice if $\ds{  \left|\frac{\gamma~(1-\delta)}{~~(1+\delta)}  \right| <1 }$. Then we obtain the domain for the existence of the zero-energy modes
\begin{equation}\label{6547uejskj2}  
- \frac{1 - \gamma}{1+\gamma}  <\delta<  1
\end{equation} 
\begin{figure}[t]\label{2678ik0}
\includegraphics[width=5.5cm]{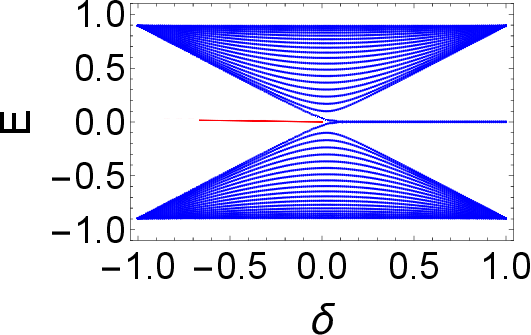}
\caption{ The plot shows the energy eigenvalues (in blue) as a function of $\ds{\delta }$ at $\ds{\gamma=0.2}$ and $\ds{N=40}$. There exists no topological zero modes in topologically trivial region $\ds{\delta<0 }$. However, zero energy quasi-stationary states (in red) exist in topologically trivial region. These zero-energy quasi-stationary states are robust to weak hopping amplitude disorder as they are for topological zero modes for $\ds{\delta>0 }$.}
\end{figure}
This solution is exact for the semi-infinite lattice while it is quasi-stationary for a sufficiently long lattice. We recover the topological zero-energy modes when $\ds{  0    <\delta   < 1    }$ and we further predict non-topological zero-energy quasi-stationary states when $\ds{- \frac{1 - \gamma}{1+\gamma}  <\delta   \leq 0    }$. In Fig. 2, we plot the energy eigenvalues as a function of $\ds{\delta}$ at $\ds{\gamma=0.2}$. As can be seen, no zero-energy modes (in blue) appear in the topologically trivial region $\ds{\delta<0}$. However, zero-energy quasi-stationary states (in red) exist in between $\ds{-0.67<\delta<0}$. They are almost stationary for a long time. For example, we numerically find that the quasi-stationary zero energy state at $\ds{\delta=-0.4}$ is almost constant up to $\ds{t=23}$. It is sufficiently long enough to realize such waves in a typical non-Hermitian experiment with waveguides \cite{fswg}.  \\
A question arises. Are these zero-energy quasi-stationary modes in the topologically trivial region robust? Let us study robustness of these non-topological zero-energy quasi-stationary modes against weak hopping amplitude disorder, which can be introduced in our system by having randomized weak forward hopping amplitudes $\ds{T_n\rightarrow T_n+\delta  T_n}$ and $\ds{\gamma \rightarrow \gamma+\delta \gamma_n}$, where $\ds{  |\delta T_n|<<1 }$ and $\ds{  |\delta \gamma_n|<<\gamma }$ are real-valued random set of constants. We numerically see that zero-energy quasi-stationary modes survive in the presence of the disorder, which shows that they are robust. The topological modes and quasi-stationary modes are robust against the same kind of disorder. We are then tempted to say that zero energy quasi-stationary modes might be topological, too. This notion needs for a further study as topological invariants introduced in the literature so far does not predict the region (\ref{6547uejskj2}). This is an open problem.

\section{Conclusion}

In this paper, we predict novel quasi-stationary solutions that are non-perturbatively different from eigenstates but preserve their forms up to a large time. Eigenvalue equations are not compatible with finding such solutions since they normally appear for infinitely long lattices. They have no analogue in Hermitian systems as they can only be seen in non-Hermitian systems. We show that very long and infinitely long lattices can have dramatically different spectra in non-Hermitian systems as opposed to Hermitian system. We discuss that the non-Hermitian skin effect is responsible for the existence of quasi-stationary solutions. We show that PBC spectrum determines the boundary of OBC spectrum in complex plane.\\
We apply our idea to a topological system and find non-topological but robust zero-energy quasi-stationary states. They are robust against the same kind of disorder as topological zero-energy eigenstates.\\
Our formalism is constructed in 1D and higher dimensional generalization of quasi-stationary solutions is straightforward. We think that they can find some applications in non-Hermitian systems. \\
This study is supported by Eskisehir Technical University Scientific Research Projects Commission under the grant no. 20ADP168.

\end{document}